\documentclass[a4paper]{llncs} 

\usepackage{a4wide}

\usepackage{tikz}
\usepackage{pgfplots}

\usepackage{lscape}
\usetikzlibrary{pgfplots.groupplots}

\usepackage{xspace}
\usepackage{tabularx}
\usepackage{url}
\usepackage[ruled,vlined, linesnumbered]{algorithm2e}
\usepackage{arydshln}

\usepackage{natbib}
\usepackage{booktabs}
\definecolor{blue2}{RGB}{50, 50, 230}
\definecolor{red2}{RGB}{230, 50, 50}

\newcounter{groupcount}
\pgfplotsset{
    draw group line/.style n args={5}{
        after end axis/.append code={
            \setcounter{groupcount}{0}
            \pgfplotstableforeachcolumnelement{#1}\of\datatable\as\cell{%
                \def\temp{#2}
                \ifx\temp\cell
                    \ifnum\thegroupcount=0
                        \stepcounter{groupcount}
                        \pgfplotstablegetelem{\pgfplotstablerow}{[index]0}\of\datatable
                        \coordinate [yshift=#4] (startgroup) at (axis cs:\pgfplotsretval,0);
                    \else
                        \pgfplotstablegetelem{\pgfplotstablerow}{[index]0}\of\datatable
                        \coordinate [yshift=#4] (endgroup) at (axis cs:\pgfplotsretval,0);
                    \fi
                \else
                    \ifnum\thegroupcount=1
                        \setcounter{groupcount}{0}
                        \draw [
                            shorten >=-#5,
                            shorten <=-#5
                        ] (startgroup) -- node [anchor=north] {#3} (endgroup);
                    \fi
                \fi
            }
            \ifnum\thegroupcount=1
                        \setcounter{groupcount}{0}
                        \draw [
                            shorten >=-#5,
                            shorten <=-#5
                        ] (startgroup) -- node [anchor=north] {#3} (endgroup);
            \fi
        }
    }
}

\newcommand{\dbg}{\emph{de Bruijn Graph}\xspace}
\newcommand{\leon}{{\sf \sc Leon}\xspace}
\newcommand{\tool}[1]{{\sf \sc #1}\xspace}
\newcommand{\fpr}{\mathcal{F}\xspace}

\begin{document}

\title{Compression of high throughput sequencing data with probabilistic de Bruijn graph}
\author{Ga\"etan Benoit\,$^{1}$, Claire Lemaitre\,$^{1}$, Dominique Lavenier\,$^{1}$ and Guillaume Rizk\,$^1$}

\institute{$^{1}$ INRIA/IRISA/GenScale, Campus de Beaulieu, 35042 Rennes cedex.}



\maketitle

\begin{abstract}

\textbf{Motivation:}
Data volumes generated by next-generation sequencing technologies is now a major concern, both for storage and transmission. This triggered the need for more efficient methods than general purpose compression tools, such as the widely used gzip.
Most reference-free tools developed for NGS data compression still use general text compression methods and fail to benefit from  algorithms already designed specifically for the analysis of NGS data.
The goal of our new method \leon is to achieve compression of DNA sequences of high throughput sequencing data, without the need of a reference genome, with techniques derived from existing assembly principles, that possibly better exploit NGS data redundancy.

\textbf{Results:}
We propose a novel method, implemented in the software \leon, for compression of DNA sequences issued from high throughput sequencing technologies. 
This is a lossless method that does not need a reference genome. Instead, a reference is built de novo from the set of reads as a probabilistic \dbg, stored in a Bloom filter. Each read is encoded as a path in this graph, storing only an anchoring kmer and a list of bifurcations indicating which path to follow in the graph.
This new method will allow to have compressed read files that also already contain its underlying \dbg, thus directly re-usable by many tools relying on this structure.
\leon achieved encoding of a \textit{C. elegans} reads set with 0.7 bits/base, outperforming  state of the art reference-free methods.


\textbf{Availability:}
Open source, under GNU affero GPL License, available for download at \url{http://gatb.inria.fr/software/leon/}

\end{abstract}

\section{Introduction}

It is now well known that data volumes generated by next-generation sequencing are a major issue. The size of the Sequence Read Archive, hosting a major part of the sequence data generated world wide, is growing very fast and now contains 2.7 petabases\footnote{\url{http://www.ncbi.nlm.nih.gov/Traces/sra/}} of DNA and RNA \citep{leinonen2010sequence}.  This is an issue both for storage and transmission, hampering collaboration between teams and long-term storage of data needed for the reproducibility of published results. Raw reads are stored in an ASCII-based text file called FASTQ containing for each read entry a read ID, a string for the sequence itself and a string of quality scores encoding estimation of accuracy for each base. It is usually compressed with the general purpose compression tool gzip\footnote{ \texttt{www.gzip.org}, Jean-Loup Gailly and Mark Adler.}. This tool is fast and largely accepted, but does not exploit specificities of sequencing data.  

Compression of sequencing data can be divided into three distinct problems: compression of read IDs, of base sequence and of quality scores. For the compression of read IDs, standard methods work very well since read ID are usually highly similar from one read to another. Compression of base sequence and quality scores are two very different problems, the former displays high redundancy across reads when depth of sequencing is high and must be lossless, whereas the latter is a very noisy signal 
on a larger alphabet size and 
where lossy compression might be acceptable.
Some publications focus solely on quality compression \citep{yu2014traversing}, and others on sequence only  \citep{janin2014beetl}.

We focus in this work on compressing  DNA sequences.
Sequence compression techniques fall into two categories: reference-based methods, such as \tool{Quip}, \tool{CRAM} and \tool{fastqz}, exploit similarities between reads and a reference genome \citep{Jones2012,Fritz2011,Bonfield2013}, whereas \emph{de novo} compression schemes in \tool{fqzcomp}, \tool{scalce}, \tool{fastqz}, \tool{DSRC}, exploit similarities between reads themselves \citep{Bonfield2013,Hach2012,deorowicz2011compression}. 
Reference based methods map reads to the genome and then only store information needed to rebuild reads : genome position and differences. While efficient, this method requires a time-consuming mapping phase to the genome, and is not possible when no close reference is known. Moreover, the reference genome is needed also for de-compressing the data and this could lead to data loss if the reference is lost or modified.
 Most \emph{de novo} methods either  (i) re-order reads to maximize similarities between consecutive reads and therefore boost compression of generic compression methods ( \tool{scalce}), or (ii) use a context-model to predict bases according to their context, followed by an arithmetic encoder (\tool{fastqz}, \tool{fqzcomp}).

As far as we know, most existing \emph{de novo} methods are improvements of text compression algorithms, but do not exploit methods and algorithms tailored for the analysis of NGS data. One exception is the method 
\tool{Quip} (in its reference-free mode), which uses sequence assembly algorithms to build a reference genome as a set of contigs and then apply a reference-based approach~\citep{Jones2012}. 
However, this method is highly dependent on the quality of the obtained contigs and appeared out-performed in a recent compression competition~\citep{Bonfield2013}. 
In this paper, we introduce \leon, a \emph{de novo}  method for lossless sequence compression using methods derived also from assembly principles; but, 
instead of building a reference as a set of sequences, the reference is represented as a light \dbg.  
Reads are then briefly represented with a kmer anchor and a list of bifurcation choices, enough to re-assemble them from the graph.

\leon achieved encoding of base sequences of \textit{E. coli} and \textit{C. elegans} reads sets with respectively 0.49 bits/base and 0.7 bits/base, corresponding to compression ratios of 16x and 11x (see details Table~\ref{tab:resultTab}).

\section{Methods}

\subsection{Overview}

Although our compression approach does not rely on a reference genome, it bears some similarities with reference-based approaches. As we do not dispose of any external data, the first step of our approach is to build \textit{de novo} a reference from the reads and then, similarly to reference-based approaches, each read is recorded as a position and a list of differences with respect to this reference. However, one major difference lies in the data structure hosting our reference : instead of a sequence or set of sequences, a \dbg is built. 

This data structure, commonly used for de novo assembly of short reads, has the advantage of representing most of the DNA information contained in the reads while getting rid of the redundancy due to sequencing coverage. The basic pieces of information in a \dbg are kmers, i.e. words of size $k$.
Once the \dbg
is built, the idea is to represent each read by a path in this graph, an anchoring node along with a list of bifurcations 
indicating which path to follow in the graph.

Since this data structure must be stored in the compressed file to reconstruct the reads, one important issue is its size. To tackle this issue, our method relies first on a good parameterization of the \dbg and secondly on its implementation as a \textit{probabilistic} data structure. The parameters are set so that the structure stores most of the important information, that is the most redundant one, while discarding the small differences, such as sequencing errors. 
Our implementation of the \dbg is based on bloom filters~\citep{kirsch_less_2006}.
Although not exact, this is very efficient to store such large data structures in main memory and then in the compressed files.

\begin{figure*}[!tpb]
 \centerline{\includegraphics[width=0.85\linewidth]{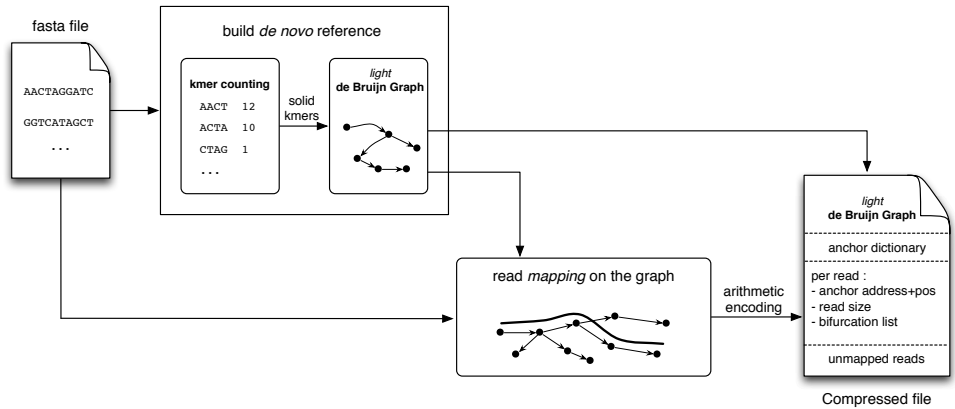}}
 \caption{\leon method overview. First, a \dbg is constructed from the reads: kmer are counted, then abundant enough kmers are inserted into a bloom filter representing a probabilistic \dbg. Reads are then mapped to this graph, and the necessary information required to rebuild the read from the graph is stored in the compressed file: an anchoring kmer and a list of bifurcations. 
 }\label{fig:leon}
 \end{figure*}

Figure~\ref{fig:leon} shows an overview of the method implemented in \leon software. First, kmers are counted and only those abundant enough 
are inserted into a bloom filter representing the \dbg. Each read is encoded by first finding its best anchoring kmer, then a walk through the graph starting from this anchor node is performed to construct the list of bifurcations followed when mapping the read to the graph. Finally, the compressed file contains the \dbg, and for each read, its anchoring kmer and a list of bifurcations encoded with an order 0 arithmetic encoder.

\subsection{Building the reference as a \dbg}

A \dbg is a directed graph where each node is a word of length $k$, called a kmer. An edge is present from node $a$ to node $b$ if the $k-1$ suffix of node $a$ is exactly the $k-1$ prefix of node $b$. 
A \dbg can be built from a set of reads by cutting each read in overlapping kmers.
Each read of size $l$ is then a path of $l-k+1$ nodes in the graph. 
In this case, the \dbg contains as many nodes as there are distinct kmers in the read dataset. 

Sequencing errors generate numerous novel distinct kmers that are present in only one or very few reads.
This increases drastically the number of nodes in the graph.
To avoid this, only kmers that are covered enough in the dataset are represented in the graph, that is kmers that have more than $T_{sol}$ (solidity threshold) occurrences in the read dataset are kept, and are hereafter called \textit{solid} kmers.

If the number of nodes is an important feature to control for the size of the data structure, the number of edges per node or the topology of the graph is also important. A given node is said to be branching if it has more than one in-going edges or more than one out-going edges. A simple path is then a path of nodes without any branching node. In order to store efficiently most of the reads, the graph should contain long simple paths such that the majority of reads will follow a simple path and no bifurcation or difference has to be stored.
This is achieved by a careful parameterization of the parameter $k$. $k$ must be large enough such that most of the kmers are unique in the target genome, but small enough such that all kmers of the genome are represented in the graph (ie. with enough occurrences in the reads).

To summarize, the reference is represented as a \dbg governed by two parameters, $k$ and $T_{sol}$. Optimal parameter values for the compression purpose depend on the dataset features, such as sequencing error rate, genome complexity and mainly the sequencing depth, as shown in Section~\ref{sec:res-features}. If both parameters are tunable by the user, the default mode of \leon does not require any user choice. The default $k$ value is $31$ and the optimal $T_{sol}$ value is inferred automatically from the analysis of the kmer counts profile.

\subsection{Probabilistic \dbg}

A traditional implementation of a  \dbg requires a lot of memory. For example, an implementation storing in a hash table for each node a kmer  and a byte containing edges, similar to the one by  \citealp{iqbal2012novo}, requires  at least $ 8 \lceil \frac{k}{32}\rceil + 1 $ bytes per node, meaning approximately $27$ GB  for a human sized genome. This is largely prohibitive for compression purposes, therefore a more lightweight implementation is required.



The notion of probabilistic \dbg was first introduced by \citealp{pell2012scaling}, and refers to a \dbg stored as a Bloom filter. It was shown that the graph nodes can be encoded with as little as 4 bits per node, with the drawback of introducing false nodes and false branching.  \citealp{Chikhi2013} then also used a bloom filter  to store the \dbg, but with an additional structure storing critical false positives, rendering the \dbg representation exact at a total cost of approximately 13 bits per node, then improved to 8 bits per node with cascading bloom filters \citep{salikhov2013using}.

For \leon compression purposes, the main issue is total graph size, and exact representation of the graph is not a major issue.
Therefore, a probabilistic \dbg was chosen for \leon, since it provides both memory-efficient representation and reasonably fast construction of the graph: the list of solid kmers are simply inserted into a bloom filter.

The \textbf{Bloom filter}~\citep{kirsch_less_2006} is a space efficient hash-based data structure, designed to test whether an element is in a set. 
It is made up of a bit array initialized with zeros, and a set of hash functions. When inserting or querying for an element, its hash values are computed yielding a set of array positions.
The insert operation corresponds to setting to 1 all these positions, whereas membership operation returns \emph{yes} if and only if all of the bits at these positions are 1.

A \emph{no} answer means the element is definitely not in the set. 
A \emph{yes} answer indicates that the element may or may not be in the set. Hence, the Bloom filter has one-sided errors. The probability of false positives increases with the number of elements inserted in the Bloom filter. 

Insertion of the graph nodes in the bloom filter is sufficient to represent the \dbg, graph edges can be inferred by querying for the existence of all 4 possible successors of a given node. 

The presence of false positive nodes in the graph is not a major issue for \leon, it only implies that additional bifurcation events may need to be stored for each read, as explained section~\ref{sec:encoding}. Therefore, there is a trade-off between the size of the bloom filter and its impact on the storage size of each read: a small bloom filter will take less space in the compressed file but will induce more storage space for each read. Since the bloom filter size is amortized across all reads, the optimal bloom filter size depends on the depth of sequencing (see Supplementary Material Section 5).

\subsection{Encoding the reads}

\label{sec:encoding}
The reference stored in the \dbg does not contain all the necessary information to retrieve a given read. The idea is to store in the compressed file the minimum information required to reconstruct a read from the graph when decompressing. The data needed is : an \emph{anchor} kmer to know where to start from in the graph, a list of \emph{bifurcation} events to tell which path to follow in the graph, and the read size to known when to stop read re-construction.

\subsubsection{Dictionary of anchors}

An \emph{anchor} kmer  is required to reconstruct a read from the graph, to know where to start graph exploration. It is the equivalent of read position in a reference genome for compression methods relying on a reference genome.

This is an important issue, a naive solution storing the raw kmer for each read would require to store for example $k$ nucleotides out of the total $l$ nucleotides of a read, representing  $k/l = 30/100$ in a typical situation, severely limiting overall compression ratio. 

\leon tackles this problem by reusing several times the same anchor kmer for different reads. Common anchor kmers are stored in a dictionary of kmers saved in the compressed file. Thus, an index in this dictionary is sufficient to encode a kmer anchor, requiring much less space than a kmer.

The selection procedure for the anchor kmer is as follows: each kmer of a read is considered as a putative anchor and queried in the dictionary of anchors. When one is found, the procedure stops and the anchor kmer is encoded as its index in the dictionary. If none is found, one \emph{suitable} anchor kmer is selected in the read then inserted in the dictionary. A \emph{suitable} kmer is a solid kmer, i.e. a kmer that is also guaranteed by design to be a graph node.

When several \emph{suitable} kmers are possible, the most abundant one is selected, it has the highest probability of being re-used by other reads.
When no \emph{suitable} kmers are found, the read cannot be mapped to the graph, it is encoded as a read without anchor, as explained section~\ref{sec:noanchor}.


\subsubsection{Bifurcation list}

The bifurcation list tells how the read is mapped to the graph, i.e. which path it follows whenever a bifurcation occurs. Since the anchoring kmer can be in the middle of the read, two bifurcation lists are needed, along with the two paths sizes. In practice, read length and anchor position is encoded, from which the two paths sizes are inferred. In the following, only the path at the right of the anchor is described, the other being symmetrical.

\begin{figure}[!tpb]
 \centerline{\includegraphics[width=0.9\linewidth]{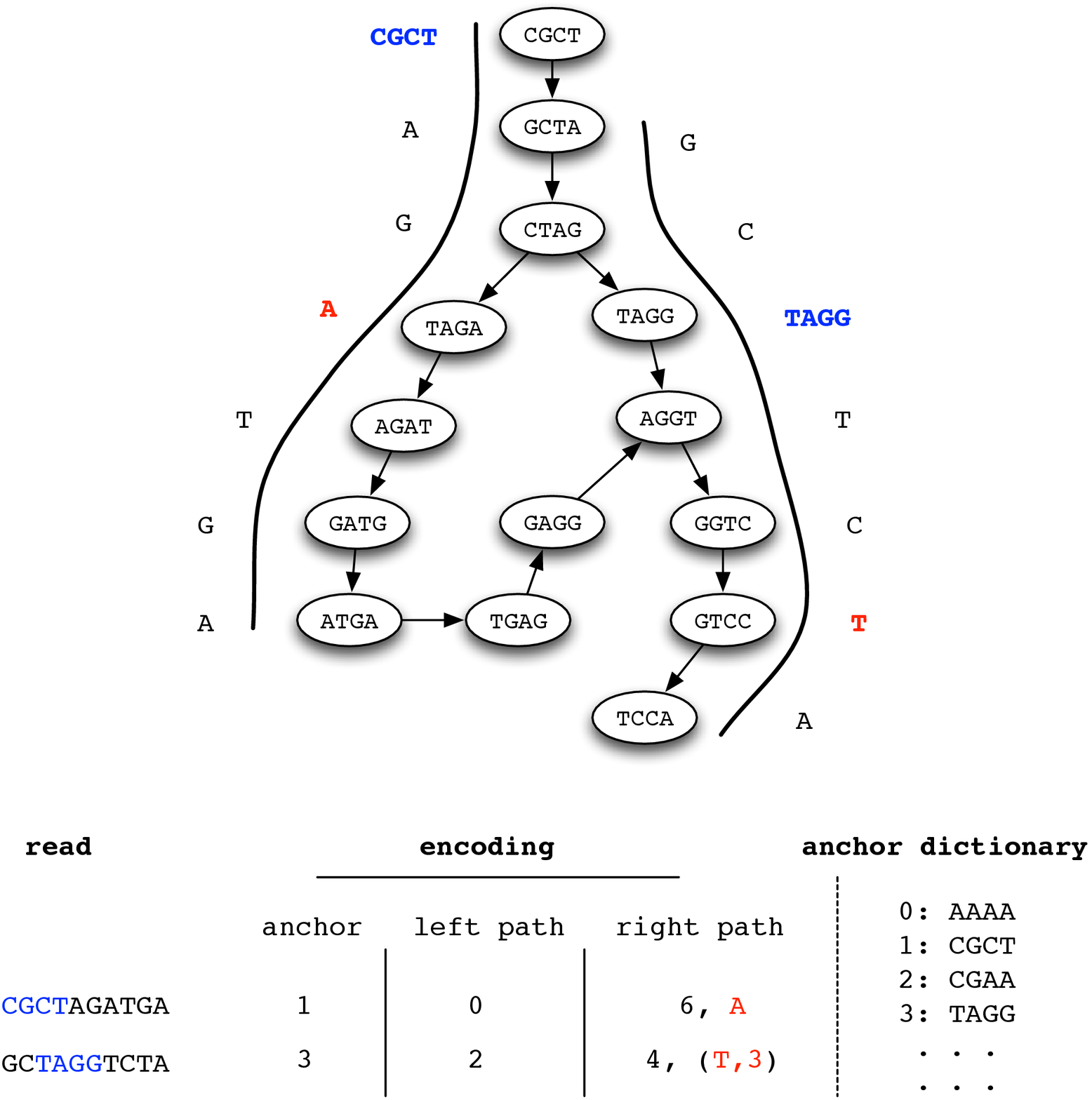}}
 \caption{Schematic description of \leon path encoding. In the upper part, the mapping of two reads to the \dbg is represented. Kmer anchors are shown in blue and bifurcations to (read at the left side) or difference from the graph (read at the right side) are highlighted in red. In the bottom part, the corresponding path encodings for these two reads are shown : the index of the kmer anchor, and for each side the path length and bifurcation list.
 }\label{fig:bif}
 \end{figure}

Starting from the anchor, the four possible kmer successors are queried in the \dbg, and compared to the following kmer in the read. If only one successor exists and is the same as the kmer in the read, this is a simple path, nothing needs to be encoded. On the contrary, whenever an ambiguity occurs, such as  several neighbors in the graph, the real nucleotide is added to the bifurcation list. It should be noted that in general, bifurcation position in the read is not required, since it is contained in the graph.
However, in the special case of a graph simple path, but different from the read, both nucleotide and read position needs to be added. This is the case for instance of a sequencing error in the read. In this case, when decompressing, the error position cannot be inferred from the graph.
Detailed construction mechanism is explained in Algorithm~\ref{LeonBifuAlgo}, and an encoding example is shown in Figure~\ref{fig:bif}.

	\begin{algorithm}[!ht]
    
    	\KwIn{read $R$, anchor position $P_{anchor}$, \dbg $G$}
		\KwOut{bifurcation list $L_b$}

 		$L_b \longleftarrow \emptyset$\\
 		
        
        Kmer $K \longleftarrow  R[P_{anchor} \;.\;. \; P_{anchor}+k-1] $\\
        \For{$i\leftarrow P_{anchor}$ \KwTo $|R|-k-1$}{
			$Nt$ = $R[i+k]$ \\
            $K_{next}$ = Suffix($K$,$k-1$) + $Nt$ \\

			$L_{succ}$ = Successors\_in\_graph($K$,$G$) \\

            \If{size of $L_{succ}$ = 1}{
				\If{$K_{next} \in L_{succ}$}{
					\tcp{nothing to do}
				}
				\Else{
                \tcp{ probable sequencing error }
					$L_b \leftarrow (Nt, i + k)$
				}
                $K$ = $L_{succ}[0]$
			}
            \Else{
            	$L_b \leftarrow Nt$\\
                $K$ = $K_{next}$\\
            }

        }
    		\caption{Bifurcation list construction.}
		\label{LeonBifuAlgo}
	\end{algorithm}

\subsubsection{Reads without anchor}
\label{sec:noanchor}
Reads that cannot be mapped to the graph are simply encoded in the file with their raw sequence of nucleotides. This happens only if all kmers of a read are not solid, i.e. if there is at least one sequencing error every $k$ nucleotides or if the reads is from a low covered region.  Therefore, this is a rare event and does not impact significantly total compression ratio in typical situations (verified experimentally, see Figure~\ref{fig:contributions}).

\subsubsection{Arithmetic coding}

All elements inserted in the compressed file (except for the bloom filter) are encoded with order 0 arithmetic coding \citep{arithm}. \leon uses an individual model for each component (read size, anchor position, bifurcation list, raw nucleotides for unmapped reads) registering symbol frequencies and assigning less bits to frequent symbols.

\subsection{Decompression}
The main difference between the decompression and the compression processes is the  reference building step that is not required during decompression since the reference \dbg is stored in the compressed file. 
The decompression process starts thus by loading in memory the \dbg and the anchor dictionary. 
Then for each read, the anchor kmer is obtained by a simple access to the dictionary of anchors.
The anchor position and the read size are then decoded to know how many nodes of the \dbg need to be explored in each direction (left and right paths).
The process to recover the read sequence in each direction starting from its anchor is similar to the one described in algorithm \ref{LeonBifuAlgo}. 
We first check in the bifurcation list if we are at a position where the nucleotide is different from any path in the \dbg (typically the case of a sequencing error). In this case, we add to the read the next nucleotide of the bifurcation list. In other cases, the successive nucleotides are obtained from the walk in the \dbg and whenever a bifurcation is encountered, the path to choose is given by decoding the next nucleotide of the bifurcation list.

\subsection{Implementation}

\subsubsection{GATB library}

The GATB library\footnote{\url{http://gatb.inria.fr/}} was used to implement \leon \citep{Drezen2014}. This library provides in particular an API for building and navigating a \dbg and its implementation, with
the Bloom filter and the constant-memory kmer counting algorithm 
introduced by \citealp{Rizk2013}. Kmer counting is the first step of \leon compression and is computationally expensive. Developing efficient and frugal kmer counting algorithms is still an open problem, with  for example recent work by \citealp{deorowicz2014kmc} showing  significant improvements. The GATB library now also implements  methods introduced by \citealp{deorowicz2014kmc}, i.e. minimizer-based kmer partitioning and $(k,x)$-mers counting.

The GATB library also provides APIs for easy sequence and kmer manipulation, as well as abstraction for system level functionalities, such as memory, disk management and multi-threading.

\subsubsection{Header compression}
To compress the sequence headers, a classic compression approach was used. A typical header string can be viewed as several fields of information separated by special characters (any character which is not a digit or alphabetic). Most of these fields are identical for all reads (for instance, the dataset name or the size of the reads). The idea is to store fixed fields only once and efficiently encode variable fields. 
A short representation of a header can be obtained using its previous header as reference. Each field of the header and its reference are compared one by one. Nothing need to be kept when fields match. 
When differences occur, either the numerical difference or the size of the longest common prefix are used to shorten the representation. The resulting short representation is encoded using an order 0 arithmetic coding.


\subsubsection{Complexity}
If we omit the kmer counting step, \leon performs compression and decompression in one single pass over the reads. For a given read, selecting the anchor and building the bifurcation list requires a number of operations that is proportional to the number of kmers in the read. 
Both compression and decompression processes have an execution time proportional to the read count times the average number of kmers per read, that is a time complexity linear with the size of the dataset.

It is important to note that decompression is faster than compression. The time consuming kmer counting step is not performed during decompression since the \dbg stored in the compressed file. 

Two main structures are maintained in main memory during compression and decompression. 
The bloom filter can use up to $G * b$ bits for storing solid kmers where $G$ is the size of the target genome and $b$ is the number of bits per solid kmers (typically $b$ is set to 12). During anchor selection, the minimum requirement is to choose a solid kmer as anchor. It means that like the bloom filter, the maximum number of anchors that can be inserted in the dictionary is $G$, the size of the genome. The important thing to notice is that the amount of memory needed by \leon is not related to the size of the input file but proportional to the size of the target genome.

\subsubsection{Parallelization}
To allow our method to fully benefit from multi-threading, reads of the input file are splitted in blocks of $n$ reads. Each block is then processed independently of the others by a given thread.
Parallelization speed-up is shown in Supplementary Material, Figure S3.


\section{Results}

\subsection{Datasets and tools}

\subsubsection{NGS datasets}

\leon performance was evaluated on several publicly available read datasets.
Main tests and comparisons were performed on whole genome sequencing (WGS) Illumina datasets with high coverage (more than 70x), from three organisms showing a large range of genome size and complexity: a bacteria \textit{E. coli} (G=5 Mbp), a nematode \textit{C. elegans} (G=100 Mbp) and a human individual (G=3 Gbp). The largest file tested is the WGS human one with 102x coverage resulting in an uncompressed fastq file size of 733 GB. To evaluate the impact of sequencing depth, these datasets were then randomly down-sampled. Additionally, other types of sequencing protocols and technologies were tested, such as RNA-seq, metagenomics, exome sequencing or Ion Torrent technology.
Detailed features and accession numbers of each dataset are given in Supplementary Table 1.

\subsubsection{Other tools and evaluation criteria}


Several compression software were run on these datasets to compare with \leon, from best state-of-the-art tools to the general purpose compressor \tool{gzip} (Supplementary Material, Table 2).

\leon and concurrent tools were compared on the following main criteria: (i) compression ratio, expressed as the original file size divided by the compressed file size, (ii) compression time, (iii) de-compression time and (iv) main memory used during compression.
Since \leon compresses only the DNA sequence and header streams, compression ratio of other tools taking only fastq format as input was computed for the sequence and header components only (see Supplementary Material for additional details and used command lines).

All tools were run on an 2.50 GHz Intel E5-2640 CPU with 12 cores 
and 192 GB of memory. All tools were set to use 8 threads.

\subsection{Impact of the parameters and \dbg false positives}

The compression ratio of \leon depends crucially on the quality of the reference that is built \textit{de novo} from the reads, the probabilistic \dbg. 

\begin{figure}[!ht]
     \centerline{\includegraphics[width=0.95\linewidth]{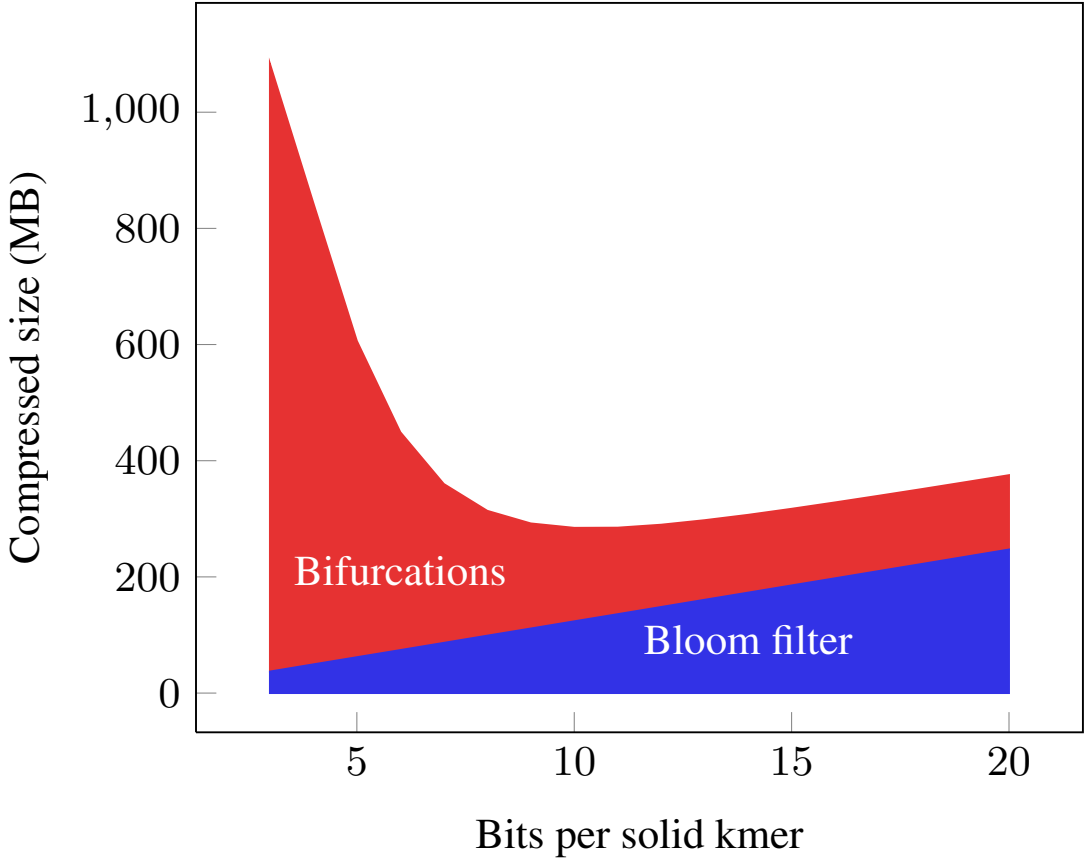}}


          
            
            
          


		\caption{Measured compressed sizes of the bifurcation list and the Bloom filter components with respect to the number of bits per solid kmer parameterized in the Bloom filter. This was obtained for the \textit{C. elegans} 70x WGS dataset.}\label{fig:nb-bits}
	\end{figure}

In order to evaluate the impact of using an approximate \dbg compared to an exact representation, the compression ratio of \leon was computed for several sizes of bloom filter expressed as a number of bits per node. The larger the bloom filter, the less false positives there are but the more space is needed to store it. Figure \ref{fig:nb-bits} shows that the optimal trade-off between the size of the structure and the cost of additional information to store per read due to false bifurcations lies around 10 bits per solid kmer, for the 70x \textit{C. elegans} dataset. Beyond 10 bits per node, the bifurcation size almost stops decreasing 
whereas the bloom filter size still increases linearly.
Importantly, these results demonstrate that correctness of \dbg is not essential for compression purposes.

The kmer size and the minimal abundance threshold (parameters $k$ and $T_{sol}$ respectively) impact also the compression ratio, as they control the number of nodes and the topology of the exact \dbg. Similarly, one can find optimal values that offer the best compromise between graph size in terms of node count and bifurcation weight. \leon compression ratio proved in fact to be robust to variations of these parameters around the optimal values. Importantly, the optimal value for the $T_{sol}$ parameter can be inferred automatically from the data, with good accuracy for WGS datasets. Concerning the $k$ parameter, it was fixed to 31 by default as a trade-off between graph size, bifurcation weight and running time (see the results on varying these parameters in Supplementary Material, Figures S1 and S2).

\subsection{Compression ratio with respect to dataset features}
\label{sec:res-features}

\begin{figure}[!ht]
         \centerline{\includegraphics[width=0.9\linewidth]{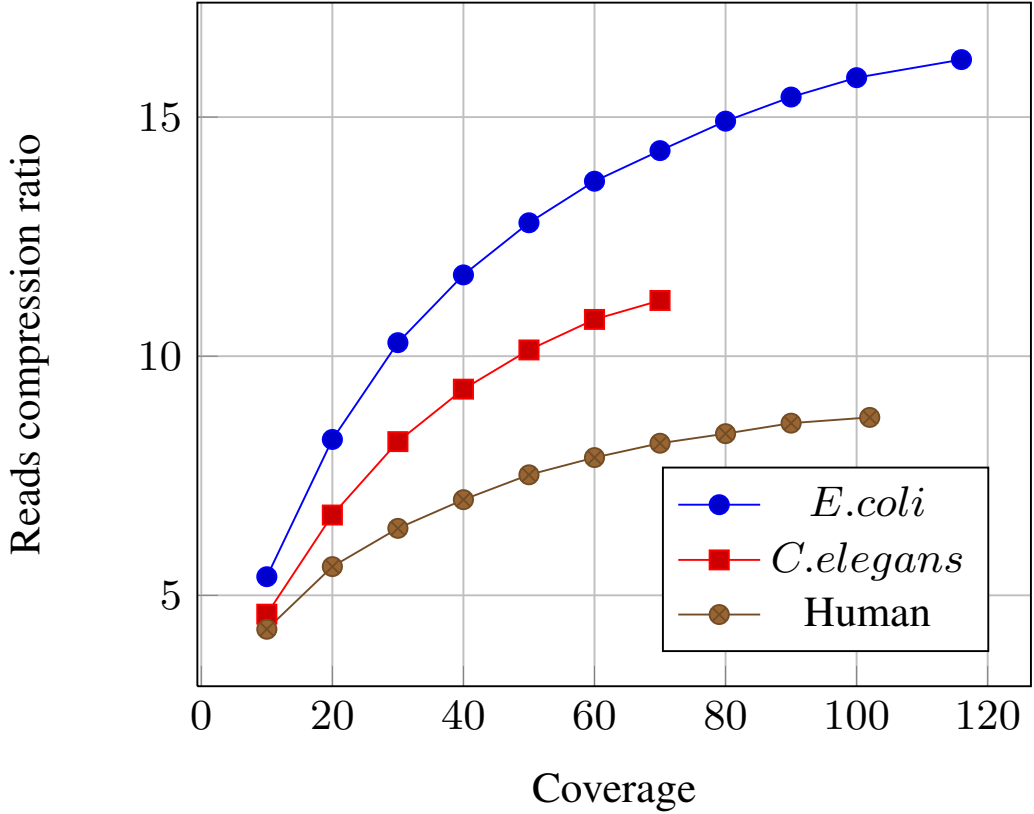}}

			

				

				
      
		\caption{Compression ratio obtained by \leon on the sequence stream with respect to the sequencing coverage of the datasets. The three WGS datasets were down-sampled to obtain lower coverage.}
        \label{fig:coverage}
	\end{figure}

\begin{figure}[!tpb]
             \centerline{\includegraphics[width=0.85\linewidth ]{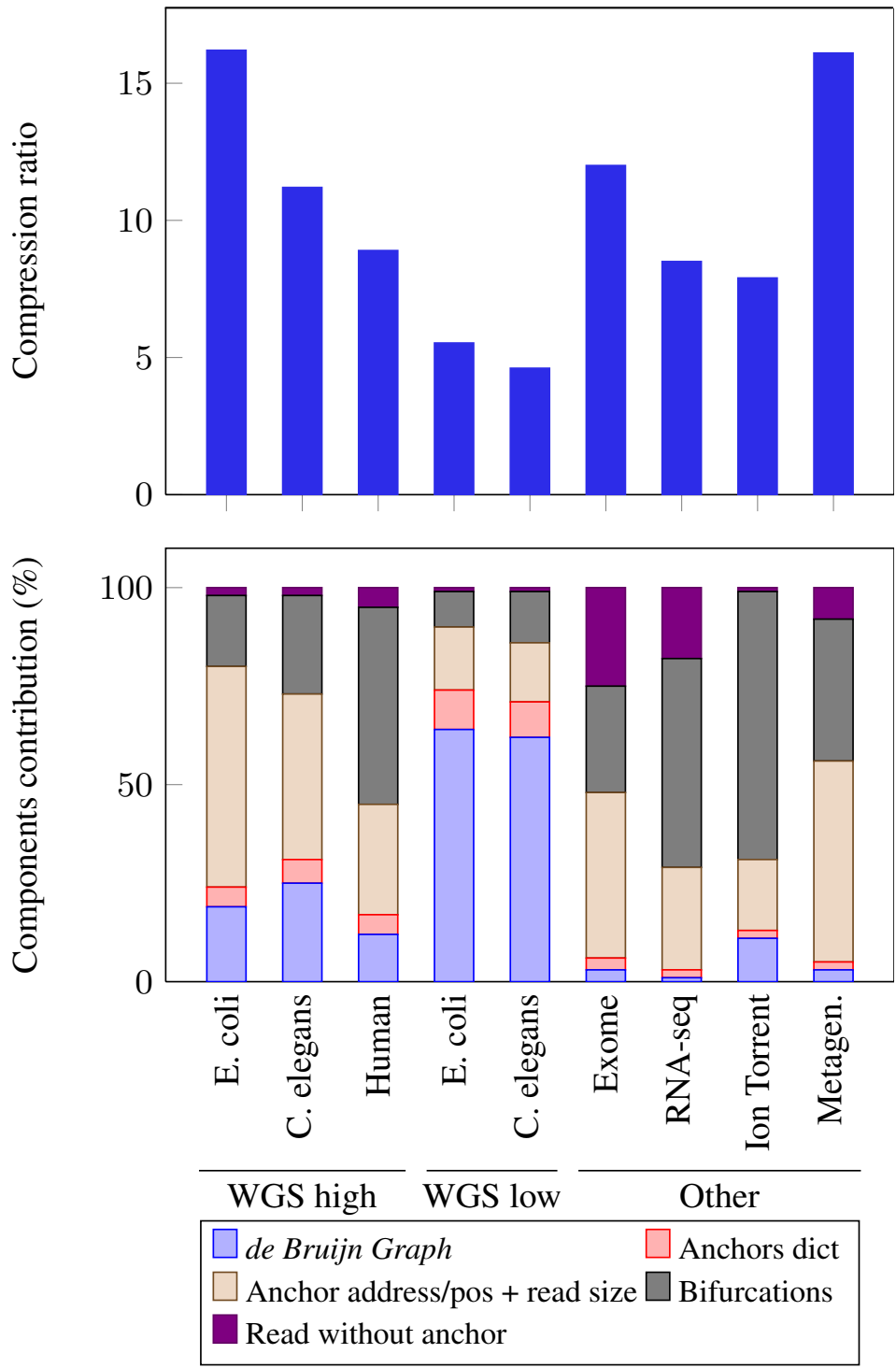}}

		\caption{Sequence compression ratio (top) and relative contribution of each component in the compressed file (bottom) for diverse datasets. WGS high means high coverage (116, 70 and 102 x respectively), WGS low means down-sampling to 10x.}\label{fig:contributions}
	\end{figure}

Figure \ref{fig:coverage} shows that the compression ratio increases with the sequencing depth. Obviously, the more redundant information is contained in the file, the more \leon can compress it. This is due to the fact that the space occupied by the Bloom filter does not depend on the sequencing depth and is rapidly negligible compared to the initial space occupied by the reads when coverage increases (see also Figure~\ref{fig:contributions}). Notably, compression factor depends also on the sequenced genome size and complexity, with better compression for the small and simple bacterial genome. In this case the \dbg contains more simple paths and bifurcation lists are smaller.

Figure~\ref{fig:contributions} shows the relative contributions of each component in the compressed file size for diverse datasets. For WGS datasets, this confirms that the relative contribution of the Bloom filter is low for high coverage datasets, but prohibitive for extremely low coverage datasets (10x). 

For other types of datasets, the relative contributions vary greatly. For instance, the exome dataset is very well compressed since the coverage is very high (more than 1000x) on a very small reference (exons representing around 1\% of the human genome). However, as the capture is noisy and some reads fall outside exons, an important part of the compressed file is taken by unmapped reads.

For the RNA-seq and ion-torrent datasets, the bifurcation component represents more than half of the compressed file size. In the ion-torrent case, this is explained by the sequencing errors that are mostly insertions and deletions, which are not well handled by the current bifurcation algorithm, contrary to substitution errors. In the RNA-seq case, due to the heterogeneous transcript abundances, some parts of the \dbg contain a high density of branchings (especially in highly transcribed genes).


\subsection{Comparison with other tools}

For high coverage WGS datasets, \leon obtained the best compression ratio for sequence and header streams combined and in particular for the sequence component 
in comparison to other compression software (see Table~\ref{tab:resultTab}). In particular, with respect to the most used tool, \tool{gzip}, \leon compressed file can be 5 times smaller than the \tool{gzip} one for high coverage datasets. In the large human dataset case, 
we can save up to 404 GB (the file size drops from 441 GB to 37 GB).  In this dataset, \tool{SCALCE} compression ratio is almost as high as \leon, but does not conserve read order.
Additional comparisons on other types of datasets are shown in Supplementary Figure S5).

Interestingly, although \tool{Quip} is similar in approach to \leon, results in terms of sequence compression ratio are much lower than \leon. This can be probably explained by a large amount of reads that could not be mapped to the assembled contigs, either because it is incomplete or too fragmented. 

As expected, compression with the reference genome allows for better compression ratio in most cases, but surprisingly this is not true for the \textit{E. coli} dataset, where \leon has a better compression ratio than \tool{Fastqz} in reference-mode.


Concerning execution time, \tool{gzip} remains by far the fastest  
software for decompressing data and require the least memory. Running time comparisons with other state-of-the-art tools are not relevant, since we can not estimate the time spent only on the sequence and header streams. We can still notice that speed values are in the same order of magnitude. For instance, decompressing the 39 MB \textit{E. coli} leon file into the 875 MB fasta file took around 20s with 8 CPU cores.

Regarding memory, contrary to other tools which use a fixed memory, memory used by \leon depends on the genome size, with less than 2 GB for a medium genome such as \textit{C. elegans}. Importantly, it remains reasonable for a human genome with 9.5GB, making \leon 
usable on desktop computers.

\setlength{\tabcolsep}{12pt}

	\begin{table}[!ht]
		\begin{tabularx}{\linewidth}{l l l l l l X }
			\hline
			\multicolumn{7}{c}{\textbf{SRR959239 - WGS \textit{E. coli} - 875 MB - 116x}} \\
			\hline
			\textbf{Prog} & \textbf{Ratio} & \textbf{Header} & \textbf{Base} & \textbf{C.Time} & \textbf{D.Time} & \textbf{Mem.}\\ 
			\hline
			Leon        & \textbf{22.7} & \textbf{45.5} & \textbf{16.2} & 24s  & 19s & 700 \\ 
			Fastqz      & 18.9 & 41.2 & 13.7 & 2m59s$^*$  & 3m34s$^*$ & 1343 \\ 
			Fqzcomp     & 16.9 & 44.4 & 11.8 & 58s$^*$  & 1m3s$^*$ & 4155 \\ 
			SCALCE      & 14   & 42.4 & 9.5  & 1m13s$^*$ & 33s$^*$ & 1844 \\ 
			Quip        & 8.9  & 40   & 5.7  & 4m43s$^*$  & 5m2s$^*$ & 780 \\ 
			gzip        & 4.4  & -    & -    & 2m21s  & 7s & 1    \\ 
            \hdashline
			Fastqz ref  & 20.8 & 41.2 & 15.4 & 3m56$^*$  &  3m6s$^*$   & 1343 \\ 
			\hline
		\end{tabularx}
		
		\vspace{0.5cm}
		
		\begin{tabularx}{\linewidth}{l l l l l l X }
			\hline
			\multicolumn{7}{c}{\textbf{SRR065390 - WGS \textit{C. elegans} - 11 GB - 70x}} \\
			\hline
			\textbf{Prog} & \textbf{Ratio} & \textbf{Header} & \textbf{Base} & \textbf{C.Time} & \textbf{D.Time} & \textbf{Mem.}\\ 
			\hline
			Leon       & \textbf{16.4} & 48   & \textbf{11.2} & 6m40s & 5m36s & 1800 \\ 
			Fastqz     & 11.1 & \textbf{61.8} & 7.3  & 37m45$^*$  & 43m47s$^*$ & 1527 \\ 
			Fqzcomp    & 11.5 & 55.6 & 7.6  & 15m30$^*$ & 19m25s$^*$ & 4208 \\ 
			SCALCE     & 12.2 & 16.5 & 10.1 & 16m21s$^*$ & 7m4s$^*$ & 5820 \\ 
			Quip       & 7.5  & 54.3 & 4.8  & 19m17s$^*$ & 20m28s$^*$ & 778 \\ 
			gzip       & 4.3  & -    &  -    & 28m6s & 1m35s & 1 \\
			\hdashline
			Fastqz ref & 25.8 & 61.8 & 18.8 & 41m43s & 30m16s & 1465 \\ 
			\hline
		\end{tabularx}
		
		\vspace{0.5cm}
		
		\begin{tabularx}{\linewidth}{l l l l l l X }
			\hline
			\multicolumn{7}{c}{\textbf{SRR345593/SRR345594 - WGS human - 441 GB - 102x}} \\
			\hline
			\textbf{Prog} & \textbf{Ratio} & \textbf{Header} & \textbf{Base} & \textbf{C.Time} & \textbf{D.Time} & \textbf{Mem.}\\ 
			\hline
			Leon      & \textbf{11.8} & \textbf{27.8} & \textbf{8.9} & 418m & 389m & 9800 \\ 
			Fastqz    & (a) &  &  &  &  & \\ 
			Fqzcomp   & 7.22 & 23.3 & 5.3 & 731m* & 875m* & 4208 \\ 
			SCALCE    & 9.5 & 11.7 & 8.7 & 1037m* & 737m* & 5342 \\ 
			Quip      & 8.6 & 17 & 4.5 & 992m* & 857m* & 780 \\ 
			gzip      & 5.4 & - & - & 1256m & 81m & 1 \\ 
			\hline
		\end{tabularx}
        
		\vspace{0.2cm}
        
		\caption{Compression features obtained for the three high coverage WGS datasets with several compression tools. The first compression ratio is given for the header and sequence streams combined, following columns indicate  ratio for each individual component.
        C.Time and D.Time denote respectively compression and decompression time, in minute, seconds. Memory is in MB.
        All tools were used without a reference genome, except for the last line of the table. Best results (without reference) are in
bold. Stars indicate that running time is obtained on Fastq files (including calculus on quality stream too). (a) Program does not support variable length sequences.}\label{tab:resultTab}
	\end{table}

\section{Discussion}

In this article, we introduced a new method for reference-free NGS data compression. While the \tool{Quip} approach is to build a \emph{de novo} reference with traditional assembly methods, we use as a \emph{de novo} reference a \dbg. This allows to skip the computationally intensive and tricky assembly step, and also allows to map more reads on the graph that would be possible on a set of \emph{de novo} built contigs.
Our approach also yields better compression ratio than context models methods such as \tool{Fastqz} or \tool{Fqzcomp}, which, in a way, also learn the underlying genome from the context. This can be explained by the larger word size used by \leon. Thanks to the probabilistic \dbg, our method is able to work with large kmers, whereas context models are limited to approximately order-14 models because of memory constraints. 

The development of an API in the GATB library to read the \leon format on-the-fly without full decompression on disk is under work and will 
facilitate usage by other tools based on GATB, that could use it as a native input format. 
Moreover, the \leon compressed file contains more information than just the raw list of reads: the included \dbg can be directly re-used by other software. 
For example, the \tool{TakeABreak} software \citep{Lemaitre2014} detecting inversions from the \dbg will be able to take as input a \leon file and save significant time from the graph construction step. In this way, \leon can be seen as more than just a compression tool, it also pre-processes data for further NGS analysis.

Further developments to enhance \leon performance and functionalities are also considered. First, if reordering reads is acceptable for the user, grouping reads with the same anchor would allow to store the anchor once for many reads and save significant space. 
Secondly, the detection of insertion and deletion errors could boost substantially the compression ratio of datasets issued from novel sequencing technologies (Ion Torrent or Pacific Bioscience).
Finally, quality compression, both lossless or lossy, is planned to allow fastq support. 

Lastly, our approach bears some similarities with error correction methods. When reads are mapped to the graph, some sequencing errors are clearly identified and saved in the file for the decompression. It could be combined with more powerful error detection algorithms to provide state-of-the art error correction, for example with the \tool{Bloocoo}\footnote{\url{http://gatb.inria.fr/software/bloocoo/}} tool already implemented with the GATB library \citep{Drezen2014}. 
It would then be straightforward to propose an option when decompressing the file, to  choose between lossless decompression mode, or with the sequencing errors corrected.

%
%


\section*{Acknowledgement}
The authors warmly thank Erwan Drezen for interesting discussions, implementation support and careful reading of the manuscript. The authors are also grateful to  Thomas Derrien, Rayan Chikhi, Raluca Uricaru and Delphine Naquin for beta-testing.
This work was supported by the ANR-12-EMMA-0019-01 GATB project.




\newpage
\onecolumn
\setcounter{section}{0}
\centerline{\bfseries\Large Supplementary material for the paper}
\bigskip
\centerline{\bfseries Compression of high throughput sequencing data with probabilistic de Bruijn graph}
\bigskip
\centerline{by}
\bigskip
\centerline{Ga\"etan Benoit, Claire Lemaitre, Dominique Lavenier, and Guillaume Rizk}

\newpage

\section{Description of the sequence datasets}

All read datasets used in the main paper are publicly available in the Sequence Read Archive (SRA) and were downloaded either from the NCBI or EBI web servers. Description of each dataset along with its SRA accession number is given in Table~\ref{tab:datasets}.

\begin{landscape}
\begin{table}[!ht]
\centering
\begin{tabular}{llllrrrrrr}
	\hline
	\textbf{SRA Accession} & \textbf{Org.} & \textbf{Type} & \textbf{Platform} & \textbf{Read size} & \textbf{Read count} & \textbf{Base count} & \textbf{Cov.} & \textbf{Fastq size} & \textbf{Fasta size}\\ 
	\hline
	SRR959239  	& \textit{E. coli} & WGS & Illumina & 98 & 5,372,832 & 526.5 Mbp & 116x & 1.4 GB & 875 MB \\  
	SRR065390 	& \textit{C. elegans} & WGS & Illumina & 100 & 67,617,092 & 6.2 Gbp & 70x & 22.6 GB & 11.4 GB\\ 
    SRR345593/SRR345594 & human & WGS & Illumina & 101 & 3,040,306,840 & 304.0 Gbp & 102x & 733 GB & 441 GB\\ 
    SRR359098/SRR359108 & human & exome & Illumina & 100 & 779,550,866 & 78.0 Gbp & $\sim 1300x$ & 203 GB & 123 GB\\ 
    SRR445718 & human & RNA-seq & Illumina & 100 & 32,943,665 & 3.3 Gbp & -- & 11 GB & 5.4 GB \\ 
    SRR857303 & \textit{E. coli} & WGS & Ion Torrent & 195 & 2,581,532 & 0.5 Gbp & 109x & 1.2 GB & 592 MB\\ 
	SRR359032 & microorg. & metagenome & Illumina & 100 & 34,690,194 & 3.5 Gbp & -- & 11 GB & 5.3 GB\\
	\hline
\end{tabular}
\caption{Read dataset description. WGS stands for Whole Genome Sequencing.}\label{tab:datasets}
\end{table}
\end{landscape}

\section{Impact of parameters $k$ and $T_{sol}$}

The kmer size and the minimal abundance threshold, ie. parameters $k$ and $T_{sol}$ respectively, impact \leon compression ratio, as they control the number of nodes and the topology of the \dbg. \leon performance was then computed for varying values of these parameters for the \textit{C. elegans} WGS dataset.

\begin{figure}[!ht]
		\centering
		\begin{tikzpicture}
			\begin{axis}[grid=major, xlabel={Solidity threshold}, ylabel={Compression ratio}, tick label style={/pgf/number format/fixed}]
			
				\addplot coordinates {
					(0, 5.564)
					(2, 10.401)
					(4, 10.954)
					(6, 11.094)
					(8, 11.162)
					(10, 11.202)
					(12, 11.219)
					(14, 11.217)
					(16, 11.188)
					(18, 11.120)
					(20, 11.007)
					(22, 10.959)
					(25, 10.44)
					(30, 9.47)
					(35, 8.28)
					(40, 7.15)
					(50, 5.64)
				};
      
			\end{axis}
		\end{tikzpicture}
		\caption{Variations of the compression ratio obtained by \leon on the \textit{C. elegans} WGS dataset for varying values of the minimal abundance threshold (parameter $T_{sol}$). \leon was run with $k=31$.}
        \label{fig:tx-tsol}
	\end{figure}
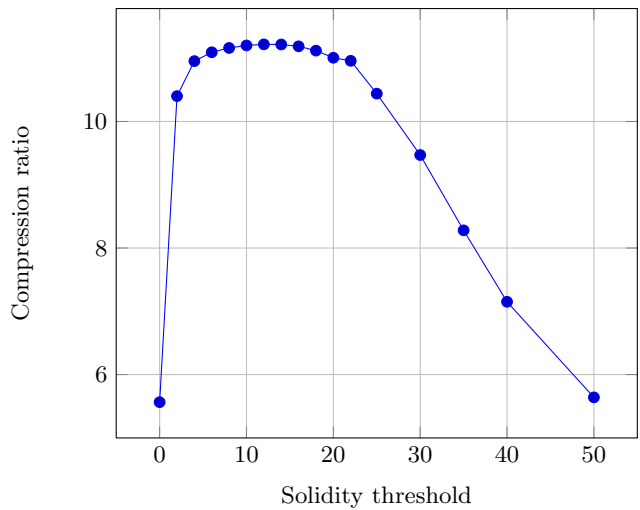

Figure~\ref{fig:tx-tsol} shows that the compression ratio is robust to variations of the $T_{sol}$ parameter around its optimal value. Importantly, the automatically inferred value, 8, 
gives a compression ratio very close to the optimal one.
However, with more extreme values the compressed file size can drastically increase. For instance, if no filtering at all of low coverage kmers is performed, the compression ratio is divided by two (from 11.2 to 5.6), demonstrating the importance of removing sequencing errors in the reference \dbg. In this case, all reads can be mapped perfectly to the graph, but the size of the graph and of the bifurcation lists are too important. Conversely, if the threshold is too high, removing too many genomic kmers, the compression ratio drops since the reference is incomplete and many more reads can not map to it.


\begin{figure}[!ht]
		\centering
		\begin{tikzpicture}
			\begin{axis}[grid=major, xlabel={kmer size}, ylabel={Compression ratio}, tick label style={/pgf/number format/fixed}]
			
				\addplot coordinates {
					(10, 4.07)
					(13, 3.97)
					(16, 6.03)
					(19, 9.7)
					(22, 10.9)
					(25, 11.1)
					(28, 11.16)
					(31, 11.16)
					(36, 10.98)
					(42, 10.95)
					(48, 10.75)
					(54, 10.30)
					(60, 9.81)
					(63, 9.6)
				};
      
			\end{axis}
		\end{tikzpicture}
		\caption{Variations of the compression ratio obtained by \leon on the \textit{C. elegans} WGS dataset for varying values of kmer size (parameter $k$). The $T_{sol}$ parameter was inferred automatically. }
        \label{fig:tx-k}
	\end{figure}
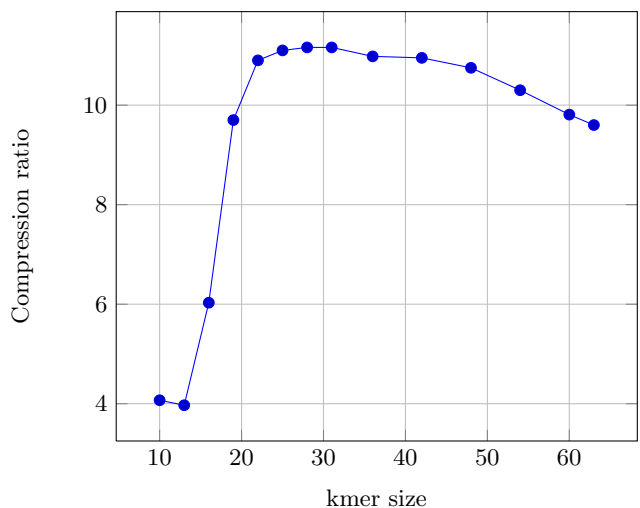

Similarly, Figure~\ref{fig:tx-k} shows that there is also an optimal value for the $k$ parameter. For small $k$ (typically $k<20$ for the \textit{C. elegans} dataset), many kmers are not unique in the genome, generating numerous branchings in the \dbg. Even if the latter is much smaller in size with small $k$, this does not compensate for the numerous bifurcations to store for each read. 
Higher $k$ value will also mean a larger dictionary of anchors and more unmapped reads.
Consequently, this parameter depends on the complexity of the genome. For large genomes with numerous repeats, larger $k$ should be preferred, 
but a trade-off must be found to compensate compression ratio with running time, since the counting step time can increase with $k$.

\section{Parallelization speed-up}

\begin{figure}[!h]
		\centering
		\begin{tikzpicture}


			\begin{axis}[grid=major, xlabel={Cores}, ylabel={Speed (Mb/s)}, tick label style={/pgf/number format/fixed}, legend style={at={(0.95,0.05)},anchor=south east}]
            
				\addplot coordinates {
					(1, 1.79)
					(2, 3.28)
					(4, 6.10)
					(8, 11.19)
					(12, 14.47)
					(16, 15.98)
					(20, 16.48)
					(24, 17.01)
				};
				\addlegendentry{Compression}


				\addplot coordinates {
					(1, 2.5)
					(2, 4.26)
					(4, 7.46)
					(8, 13.84)
					(12, 19.94)
					(16, 23.82)
					(20, 27.54)
					(24, 31.44)
				};
				\addlegendentry{Decompression}
      
			\end{axis}

		\end{tikzpicture}
        
		\caption{Speed of compression and decompression with respect to the number of used CPU threads, for the 70x WGS \textit{C. elegans} dataset.}
        \label{fig:speedup}
	\end{figure}
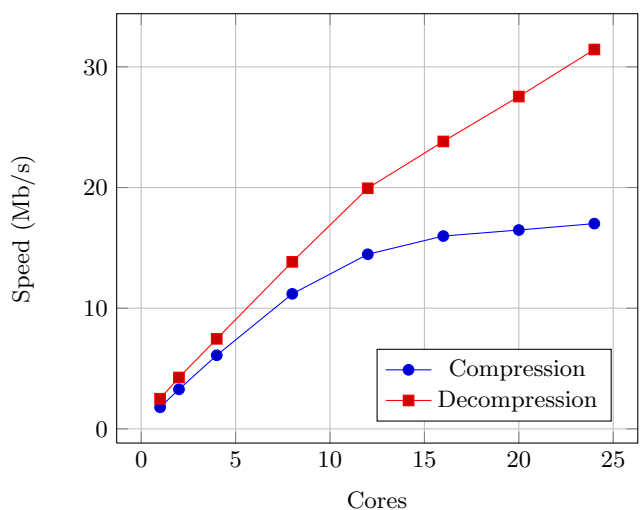

Figure~\ref{fig:speedup} shows \leon execution time using 1 to 24 threads. The platform used is a 2.50 GHz Intel E5-2640 CPU with 12 cores ( 24 logical cores with hyper-threading). It should be noted that \leon seems to benefit highly from Intel hyper-threading. We suspect this is the case because \leon major operations are queries inside a bloom filter, which induce many memory cache misses. Memory cache misses induce latency, usually well hidden by hyper-threading.

\section{Comparisons with other compression software}

\subsection{Data formats and command line arguments}

Software used to compare to \leon are described in Table~\ref{tab:software}.

\begin{table}[!ht]
\centering
\begin{tabular}{lllll}
	\hline
	\textbf{Sotware} & \textbf{Ref.} & \textbf{Version} & \textbf{Mode} & \textbf{Command line}\\ 
	\hline
	\leon  	& -- & 0.2.1 & default & \texttt{leon -file file.fasta -c -nb-cores 8} \\ 
	\tool{Fastqz}   & \cite{Bonfield2013} & 1.5 & slow & \texttt{fastqz c file.fastq output}\\ 
    \tool{Fastqz}   & \cite{Bonfield2013} & 1.5  & with reference & \texttt{fastqz c file.fastq output ref.fapack}\\ 
	\tool{Fqzcomp}  & \cite{Bonfield2013}  & 4.6  & slow & \texttt{fqz\_comp -n2 -q1 -s8+ -b file.fastq} \\ 
	\tool{SCALCE} & \cite{Hach2012}  & 2.7  & slow & \texttt{scalce -c bz -T 8 -A file.fastq}  \\ 
    \tool{Quip} & \cite{Jones2012}  & 1.1.8  & assembly & \texttt{quip -i fastq -a file.fastq}  \\ 
	\tool{gzip}   & \cite{gzip} & 1.3.12  & & \texttt{gzip file.fasta} \\ 
	\hline
\end{tabular}
\caption{Software description and used command lines.}\label{tab:software}
\end{table}

Since \leon is focused on fasta file only (compression of DNA sequence and header), compression ratio of other tools taking only fastq format as input was obtained for the sequence and header components only, thanks to additional information given by the tools. It was not possible to obtain these decomposition for the software  \tool{DSRC}, therefore it was not tested~\cite{deorowicz2011compression}.

Tools that compress also the quality values were parameterized when possible, so that to minimize the time spent on quality values. 


\subsection{Additional comparison results}

\begin{figure}[!ht]
		\centering
		\begin{tikzpicture}

            \begin{groupplot}[
                group style={
                    group name=my plots,
                    group size=3 by 2,
                    ylabels at=edge left,
                    horizontal sep=10pt,
                    vertical sep=40pt
                },grid=none, height=5cm, xbar interval,
                symbolic y coords={quip, scalce, fqzcomp, fastqz, leon, none}, ytick=data, ymajorgrids=false, xtick pos=left, ytick pos=bottom, y tick label style={major tick length=0pt}, xmin=0, xmax=25, tick label style={/pgf/number format/fixed}, xtick={0,10,20}
            ]

				\nextgroupplot[title={WGS high}, width=6cm]
                \addplot coordinates {
                    (4.8,quip)
                    (10.1,scalce)
                    (7.6,fqzcomp)
                    (7.3,fastqz)
                	(11.2,leon)
                    (0.0,none)};
      
				\nextgroupplot[title={WGS low}, width=6cm, yticklabels={,,}]
                \addplot coordinates {
                    (4.59,quip)
                    (5.86,scalce)
                    (5.52,fqzcomp)
                    (6.23,fastqz)
                	(4.61,leon)
                    (0,none)};
                    
				\nextgroupplot[title={Exome}, width=6cm, yticklabels={,,}]
                \addplot coordinates {
                    (4.82,quip)
                    (16.8,scalce)
                    (8.77,fqzcomp)
                    (7.25,fastqz)
                	(12,leon)
                    (0,none)};
      
				\nextgroupplot[title={RNA-seq}, width=6cm]
                \addplot coordinates {
                    (8.67,quip)
                    (11.45,scalce)
                    (11,fqzcomp)
                    (10.12,fastqz)
                	(9.2,leon)
                    (0,none)};
      
				\nextgroupplot[title={Ion torrent}, width=6cm, yticklabels={,,}, xlabel={Compression ratio}]
                \addplot coordinates {
                    (5.49,quip)
                    (6.69,scalce)
                    (11.2,fqzcomp)
                    (0,fastqz)
                	(7.9,leon)
                    (0,none)};
      
				\nextgroupplot[title={Metagenomic}, width=6cm, yticklabels={,,}]
                \addplot coordinates {
                    (15.7,quip)
                    (19.59,scalce)
                    (15.3,fqzcomp)
                    (14.02,fastqz)
                	(16.4,leon)
                    (0,none)};
                    
			\end{groupplot}

		\end{tikzpicture}
        
		\caption{Comparison of the sequence compression ratio between \textit{de novo} compression software for diverse datasets. \textit{WGS high} refers to the 70x WGS \textit{C.elegans} dataset, \textit{WGS low} is the same dataset down-sampled to 10x coverage. For the RNA-seq and exome datasets, \leon was run with parameter $T_{sol}=20$. Fastqz doesn't work on the Ion torrent dataset because it contains variable length reads.}
        \label{fig:comp-special}
	\end{figure}
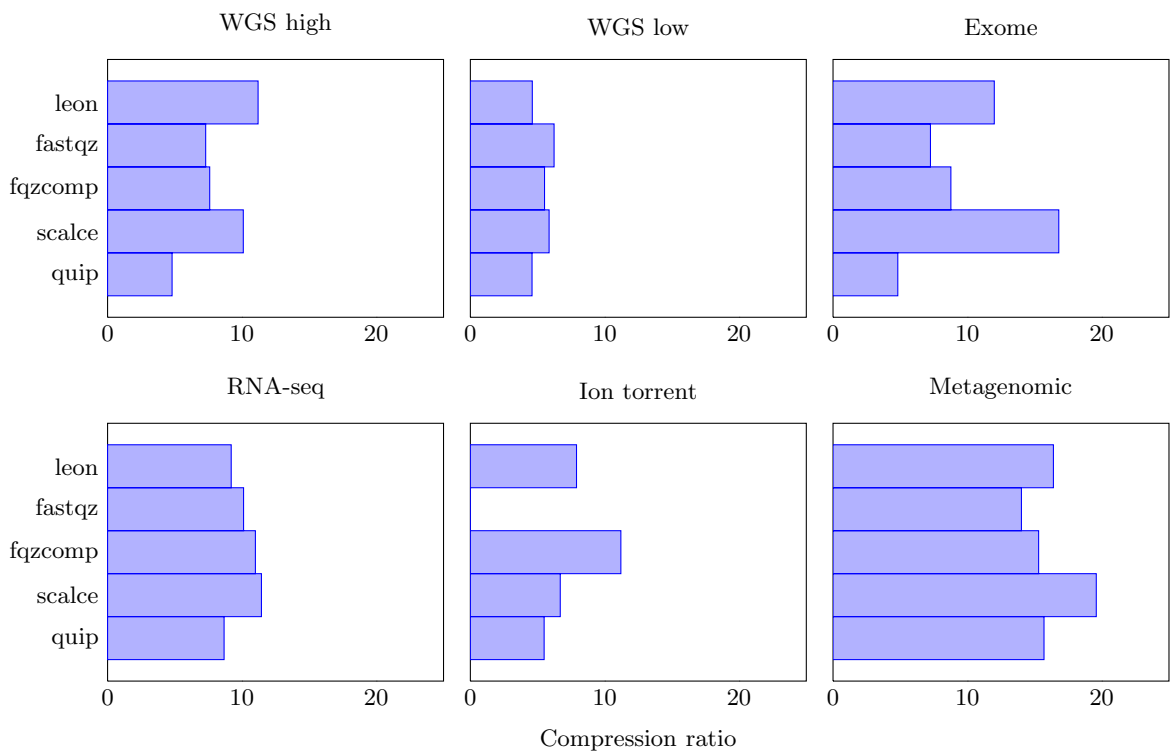

Comparisons of the sequence compression ratio between all reference-free sequence compression software for special datasets (RNA-seq, Ion-Torrent, etc.) are shown in Figure~\ref{fig:comp-special}.

\section{Theoretical estimation of the optimal bloom filter size}

\leon uses a probabilistic \dbg, i.e. all kmer nodes of the graph are inserted in a bloom filter. The false positive rate of the bloom filter will induce false branching in the graph, meaning extra bifurcation events that will need to be stored in the compressed file.
Therefore, an optimal trade-off needs to be found: a large bloom filter will take more space in the file but will save space for read storage (and conversely).

The false positive rate $\fpr$ of a bloom filter can be approximated by  $\fpr = f^r$  with $f= 0.5^{\log 2} \approx 0.618$ and $r = \frac{\mbox{bloom size} }{\mbox{nb elements inserted}}$ the number of bits per element inserted in the bloom filter~\cite{kirsch_less_2006}.

With $\mathcal{G}$ the size of the target sequenced genome (i.e. approximately the number of nodes in the \dbg) and $\mathcal{D}$ 
the average kmer abundance (somehow related to the depth of sequencing), 
the total size of the bloom filter and the false positive bifurcation events stored in the file can be approximated by:


$$
\underbrace{r \cdot \mathcal{G}}_\textrm{Bloom filter}  +   \underbrace{3\cdot 2 \cdot   \mathcal{G} \mathcal{D}  f^r }_\textrm{false bifurcations} \mbox{    bits}
$$

In average, most of graph nodes are in simple paths, hence 3 possible edges out of each node are likely to produce false bifurcations, and each bifurcation is stored with approximately 2 bits (through the arithmetic coder).

This total size is minimized with :

$$
r = \log (\frac{-1}{6\cdot \mathcal{D}  \log f}) / \log f
$$

This yields $r= 10.3 $ bits for $\mathcal{D} = 50$, very close to the experimentally measured optimal size for the SRR065390 \emph{C.elegans} dataset (70x total coverage $\approx$ 50x kmer coverage for $k=31$ and read length 100).

\bibliographystyle{natbib}

\bibliography{compression_biblio,supplementary_biblio}

\begin{thebibliography}{}

\bibitem[Bonfield and Mahoney(2013)Bonfield and Mahoney]{Bonfield2013}
Bonfield, J.~K. and Mahoney, M.~V. (2013).
\newblock Compression of fastq and sam format sequencing data.
\newblock {\em PLoS One\/}, {\bf 8}(3), e59190.

\bibitem[Chikhi and Rizk(2013)Chikhi and Rizk]{Chikhi2013}
Chikhi, R. and Rizk, G. (2013).
\newblock Space-efficient and exact de bruijn graph representation based on a
  bloom filter.
\newblock {\em Algorithms Mol Biol\/}, {\bf 8}(1), 22.

\bibitem[Deorowicz and Grabowski(2011)Deorowicz and
  Grabowski]{deorowicz2011compression}
Deorowicz, S. and Grabowski, S. (2011).
\newblock Compression of dna sequence reads in fastq format.
\newblock {\em Bioinformatics\/}, {\bf 27}(6), 860--862.

\bibitem[Deorowicz {\em et~al.}(2014)Deorowicz, Kokot, Grabowski, and
  Debudaj-Grabysz]{deorowicz2014kmc}
Deorowicz, S., Kokot, M., Grabowski, S., and Debudaj-Grabysz, A. (2014).
\newblock Kmc 2: Fast and resource-frugal $ k $-mer counting.
\newblock {\em arXiv preprint arXiv:1407.1507\/}.

\bibitem[Drezen {\em et~al.}(2014)Drezen, Rizk, Chikhi, Deltel, Lemaitre,
  Peterlongo, and Lavenier]{Drezen2014}
Drezen, E., Rizk, G., Chikhi, R., Deltel, C., Lemaitre, C., Peterlongo, P., and
  Lavenier, D. (2014).
\newblock Gatb: Genome assembly \& analysis tool box.
\newblock {\em Bioinformatics\/}.

\bibitem[Fritz {\em et~al.}(2011)Fritz, Leinonen, Cochrane, and
  Birney]{Fritz2011}
Fritz, M. H.-Y., Leinonen, R., Cochrane, G., and Birney, E. (2011).
\newblock Efficient storage of high throughput sequencing data using
  reference-based compression.
\newblock {\em Genome Res\/}, {\bf 21}, 734--740.

\bibitem[Hach {\em et~al.}(2012)Hach, Numanagic, Alkan, and Sahinalp]{Hach2012}
Hach, F., Numanagic, I., Alkan, C., and Sahinalp, S.~C. (2012).
\newblock Scalce: boosting sequence compression algorithms using locally
  consistent encoding.
\newblock {\em Bioinformatics\/}, {\bf 28}(23), 3051--3057.

\bibitem[Iqbal {\em et~al.}(2012)Iqbal, Caccamo, Turner, Flicek, and
  McVean]{iqbal2012novo}
Iqbal, Z., Caccamo, M., Turner, I., Flicek, P., and McVean, G. (2012).
\newblock De novo assembly and genotyping of variants using colored de bruijn
  graphs.
\newblock {\em Nature genetics\/}, {\bf 44}(2), 226--232.

\bibitem[Janin {\em et~al.}(2014)Janin, Schulz-Trieglaff, and
  Cox]{janin2014beetl}
Janin, L., Schulz-Trieglaff, O., and Cox, A.~J. (2014).
\newblock Beetl-fastq: a searchable compressed archive for dna reads.
\newblock {\em Bioinformatics\/}, page btu387.

\bibitem[Jones {\em et~al.}(2012)Jones, Ruzzo, Peng, and Katze]{Jones2012}
Jones, D.~C., Ruzzo, W.~L., Peng, X., and Katze, M.~G. (2012).
\newblock Compression of next-generation sequencing reads aided by highly
  efficient de novo assembly.
\newblock {\em Nucleic Acids Res\/}, {\bf 40}(22), e171.

\bibitem[Kirsch and Mitzenmacher(2006)Kirsch and
  Mitzenmacher]{kirsch_less_2006}
Kirsch, A. and Mitzenmacher, M. (2006).
\newblock Less hashing, same performance: Building a better bloom filter.
\newblock {\em {Algorithms–ESA} 2006\/}, pages 456--467.

\bibitem[Leinonen {\em et~al.}(2010)Leinonen, Sugawara, and
  Shumway]{leinonen2010sequence}
Leinonen, R., Sugawara, H., and Shumway, M. (2010).
\newblock The sequence read archive.
\newblock {\em Nucleic acids research\/}, page gkq1019.

\bibitem[Lemaitre {\em et~al.}(2014)Lemaitre, Ciortuz, and
  Peterlongo]{Lemaitre2014}
Lemaitre, C., Ciortuz, L., and Peterlongo, P. (2014).
\newblock Mapping-free and assembly-free discovery of inversion breakpoints
  from raw ngs reads.
\newblock In A.-H. Dediu, C.~Martín-Vide, and B.~Truthe, editors, {\em
  Algorithms for Computational Biology\/}, volume 8542 of {\em Lecture Notes in
  Computer Science\/}, pages 119--130. Springer International Publishing.

\bibitem[P. and J.L.(1996)P. and J.L.]{gzip}
P., D. and J.L., G. (1996).
\newblock Zlib compressed data format specification version 3.3.
\newblock {\em RFC 1950\/}.

\bibitem[Pell {\em et~al.}(2012)Pell, Hintze, Canino-Koning, Howe, Tiedje, and
  Brown]{pell2012scaling}
Pell, J., Hintze, A., Canino-Koning, R., Howe, A., Tiedje, J.~M., and Brown,
  C.~T. (2012).
\newblock Scaling metagenome sequence assembly with probabilistic de bruijn
  graphs.
\newblock {\em Proceedings of the National Academy of Sciences\/}, {\bf
  109}(33), 13272--13277.

\bibitem[Rizk {\em et~al.}(2013)Rizk, Lavenier, and Chikhi]{Rizk2013}
Rizk, G., Lavenier, D., and Chikhi, R. (2013).
\newblock Dsk: k-mer counting with very low memory usage.
\newblock {\em Bioinformatics\/}, {\bf 29}(5), 652--653.

\bibitem[Salikhov {\em et~al.}(2013)Salikhov, Sacomoto, and
  Kucherov]{salikhov2013using}
Salikhov, K., Sacomoto, G., and Kucherov, G. (2013).
\newblock Using cascading bloom filters to improve the memory usage for de
  brujin graphs.
\newblock In {\em Algorithms in Bioinformatics\/}, pages 364--376. Springer.

\bibitem[Witten {\em et~al.}(1987)Witten, Neal, and Cleary]{arithm}
Witten, I., Neal, R., and Cleary, J. (1987).
\newblock Arithmetic coding for data compression.
\newblock {\em Communications of the ACM\/}.

\bibitem[Yu {\em et~al.}(2014)Yu, Yorukoglu, and Berger]{yu2014traversing}
Yu, Y.~W., Yorukoglu, D., and Berger, B. (2014).
\newblock Traversing the k-mer landscape of ngs read datasets for quality score
  sparsification.
\newblock In {\em Research in Computational Molecular Biology\/}, pages
  385--399. Springer.

\end{thebibliography}

\end{document}